\documentclass[conference]{IEEEtran}
\IEEEoverridecommandlockouts

\usepackage{cite}
\usepackage{amsmath,amssymb,amsfonts}
\usepackage{graphicx}
\usepackage{textcomp}
\usepackage{xcolor}
\usepackage{algorithm}
\usepackage{algpseudocode}
\usepackage{listings}
\usepackage{subcaption}
\lstdefinestyle{log}{
  basicstyle=\ttfamily\footnotesize,
  columns=fullflexible,
  frame=single,
  breaklines=true,
  breakatwhitespace=true,
  keepspaces=true,
  numbers=left,
  numbersep=6pt,
  xleftmargin=1em,
  showstringspaces=false,
  captionpos=b
}

\usepackage{booktabs}
\usepackage{tabularx}
\usepackage{minted} 

\newcommand{\Marcos}[1]{\textcolor{cyan}{Marcos: {#1}}}

\def\BibTeX{{\rm B\kern-.05em{\sc i\kern-.025em b}\kern-.08em
    T\kern-.1667em\lower.7ex\hbox{E}\kern-.125emX}}
\begin{document}

\title{Enabling Real-Time Phase Control in Traffic Signal Hardware-in-the-Loop Simulation\\
\thanks{This material is based upon work supported by the National Science Foundation under Grant No. CNS-2434400 (Sprinkle, Work), and the Transportation Network Growth Opportunity (TNGO) initiative of the Tennessee Department of Economic and Community Development. Any opinions, findings, and conclusions or recommendations expressed in this material are those of the author(s) and do not necessarily reflect the views of the National Science Foundation or the Tennessee Department of Economic and Community Development.}
\thanks{© 2026 IEEE.  Personal use of this material is permitted.  Permission from IEEE must be obtained for all other uses, in any current or future media, including reprinting/republishing this material for advertising or promotional purposes, creating new collective works, for resale or redistribution to servers or lists, or reuse of any copyrighted component of this work in other works.}
}

\author{\IEEEauthorblockN{Zhiyao Zhang, Gergely Zach\'{a}r, William Barbour, Matt Bunting, \\Marcos Quiñones-Grueiro, Jonathan Sprinkle, Dan Work}
\IEEEauthorblockA{\textit{Institute for Software Integrated Systems} \\
\textit{Vanderbilt University}\\
Nashville, U.S. \\
\{zhiyao.zhang, gergely.zachar, william.w.barbour, matthew.r.bunting, \\marcos.quinones.grueiro, jonathan.sprinkle, dan.work\}@vanderbilt.edu}
}
\vspace{0.4in}
\maketitle

\begin{abstract}

Advanced Traffic Signal Control (TSC) algorithms require real-time phase control, yet existing Hardware-in-the-Loop Simulation (HILS) testbeds only support pre-programmed timing plans. In this paper, we present the first HILS testbed for real-time phase control. We develop a novel middleware architecture that translates dynamic phase actions (selection, switch, and duration) into commands for NTCIP-compliant commercial hardware controllers. This middleware manages phase transitions, synchronizes signal states, and handles errors without interrupting the hardware's internal operations. Experimental validation demonstrates that the system executes real-time phase commands, handles system conflicts, and achieves a low system internal latency at sub-millisecond on average.
\end{abstract}


\section{Introduction}
\begin{figure}
    \centering
    \includegraphics[width=1\linewidth]{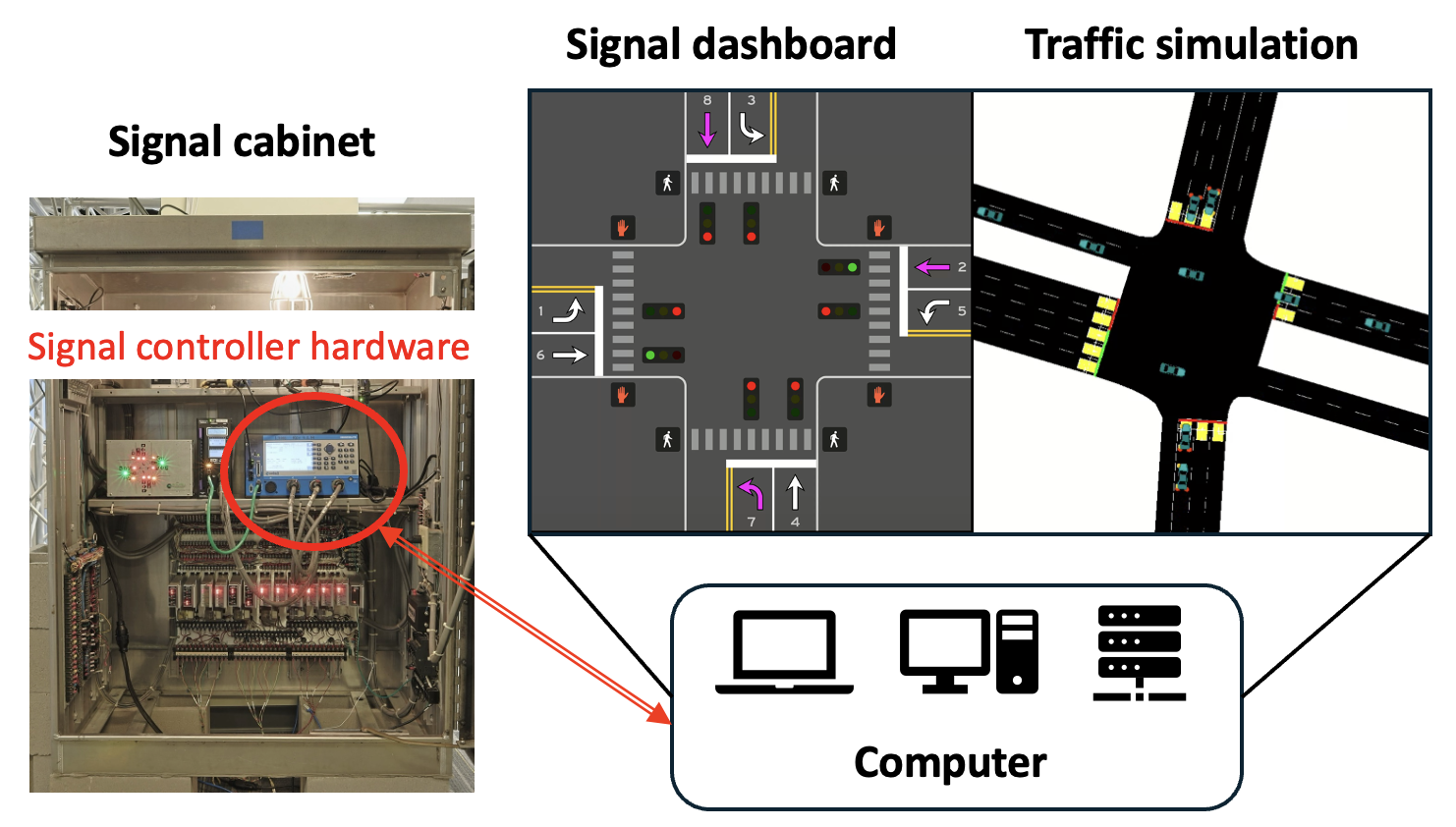}
    \caption{Our Hardware-in-the-Loop Simulation testbed enables running traffic signal control algorithms on any computational device (bottom, such as laptop, desktop, remote server) connected to the signal controller hardware (upper left, installed in signal cabinet) with standardized communication (red double-line arrow). Users can observe the signal status on a dashboard (middle) and the traffic simulation progress (right) during simulation.}
    \label{fig:cabinet}
    \vspace{-7pt}
\end{figure}
Traffic signal control (TSC) is a fundamental component of urban traffic management that directly impacts mobility, safety, and emissions at signalized intersections. Broadly, TSC operates in two paradigms. The first is \textit{plan-based control}, where signal timing parameters such as cycle length, phase splits, and offsets are pre-computed offline and uploaded to the controller\footnote{We refer controller to the signal controller hardware, and refer control algorithm/agent to the software entity that generates control actions.} as fixed or time-of-day plans. Systems such as TRANSYT \cite{robertson1969transyt}, SCOOT \cite{hunt1981scoot}, and SCATS \cite{lowrie1990scats} fall within this category, adjusting timing plans either offline or through progressive adaptive mechanisms. The second paradigm is \textit{real-time phase control}, where the green-light phases (i.e., traffic movements) are determined at each decision step based on the current traffic state. This paradigm includes both model-based methods, such as max-pressure control \cite{varaiya2013max} and its variants \cite{levin2020max, mercader2020max}, and model-free methods, most notably deep reinforcement learning (RL) approaches \cite{wei2021recent, noaeen2022reinforcement, zhao2024survey}. 
Common action formulations for real-time phase control include \textit{phase selection} (choosing the next phase to activate), \textit{phase switch} (deciding whether to keep or change the current phase), and \textit{phase duration} (setting the green time for the current or upcoming phase).

Despite the growing body of work on real-time TSC algorithms, the vast majority of these methods have been developed and evaluated exclusively in software simulation environments with manually implemented signal transition processes \cite{ault2021reinforcement,mei2024libsignal,tang2019cityflow}. 
However, signal controller hardware is by default designed for operating signal timing plans, and how to enable real-time phase control on hardware still remains an open problem. 

Hardware-in-the-Loop Simulation (HILS) has long been recognized as a higher-fidelity alternative to software-only simulation for validating signal control strategies and performance \cite{bullock2004hardware, engelbrecht2001using}, which is particularly beneficial and critical for real-time phase control as it much more frequently interacts with the hardware controller. Existing literature on HILS for TSC widely explored optimizing and running signal timing plans. Earlier explorations including \cite{engelbrecht1999development, engelbrecht2001using} demonstrated the feasibility of HILS for traffic performance evaluation and signal timing validation. Byene et al. \cite{byrne2005using} further evaluated the effect of stop location in transit signal priority performance with an HILS. Li et al. \cite{li2016new} proposed a new HILS architecture to split out the simulation process and controller logic. This modularized design is inherited in our work. Wang et al. \cite{wang2021virtual} further proposed a virtual interface over the physical hardware to replace the physical controller interface device requirement, which reduces the cost and complexity to build the HILS at large scale. However, these studies assume the signal controllers run timing plans by default, and there is yet no realization of HILS for real-time phase control.

Enabling real-time phase control in HILS introduces several technical challenges that do not arise in plan-based setups. First, \textit{real-time communication} is required: the control algorithm must receive the current signal state, compute a phase command, and transmit it to the controller within a tight time window (on the order of seconds or sub-seconds). Second, \textit{system reliability and error handling} need to be addressed: during real-time operation, a variety of errors and/or conflicts may occur, e.g., communication timeouts, simulation time drifts, phase command conflicts. The HILS must detect and handle such errors and conflicts internally without interrupting the hardware controller.
This leads to our research question: \textit{How to design a HILS testbed that supports real-time phase control?}

The main contribution of this work is a HILS for TSC with real-time phase control capability. Specifically:
\begin{enumerate}
    \item We develop a Python-based HILS that for the first time enables real-time phase control for TSC. The system is validated on a commercial controller with NTCIP 1202 v02. It supports three commonly used real-time control formulations, namely phase selection, phase switch, and phase duration, which encompass both acyclic and cyclic action designs. 
    
    \item We propose a middleware architecture with a manager and a communication layer. The manager commands new phases and retrieves signal states to and from the controller with the capability to handle errors and conflicts internally. The communication layer realizes communication to the controller in a NTCIP-compliant manner.
    
    \item We record the trajectory of the middleware responses to phase commands and demonstrate the full process from generating a phase command to verifying the transition termination. We also show the internal latency of action executions from the action generation to sending phase commands. The mean latency is on the order of one millisecond with negligible impact to the system performance.
    
\end{enumerate}

\section{Preliminaries}
We introduce the preliminaries about the terms and concepts related to this work.

\noindent\textbf{Phases and phase pairs:}
A phase $p\in\mathcal{P}$ is an elementary signal indication pattern that serves a specific traffic movement (or a compatible set of movements) with right-of-way, typically a green indication for those movements while conflicting movements are held red. Two compatible phases that can run concurrently form a phase pair.

\begin{figure}
    \centering
    \includegraphics[width=1\linewidth]{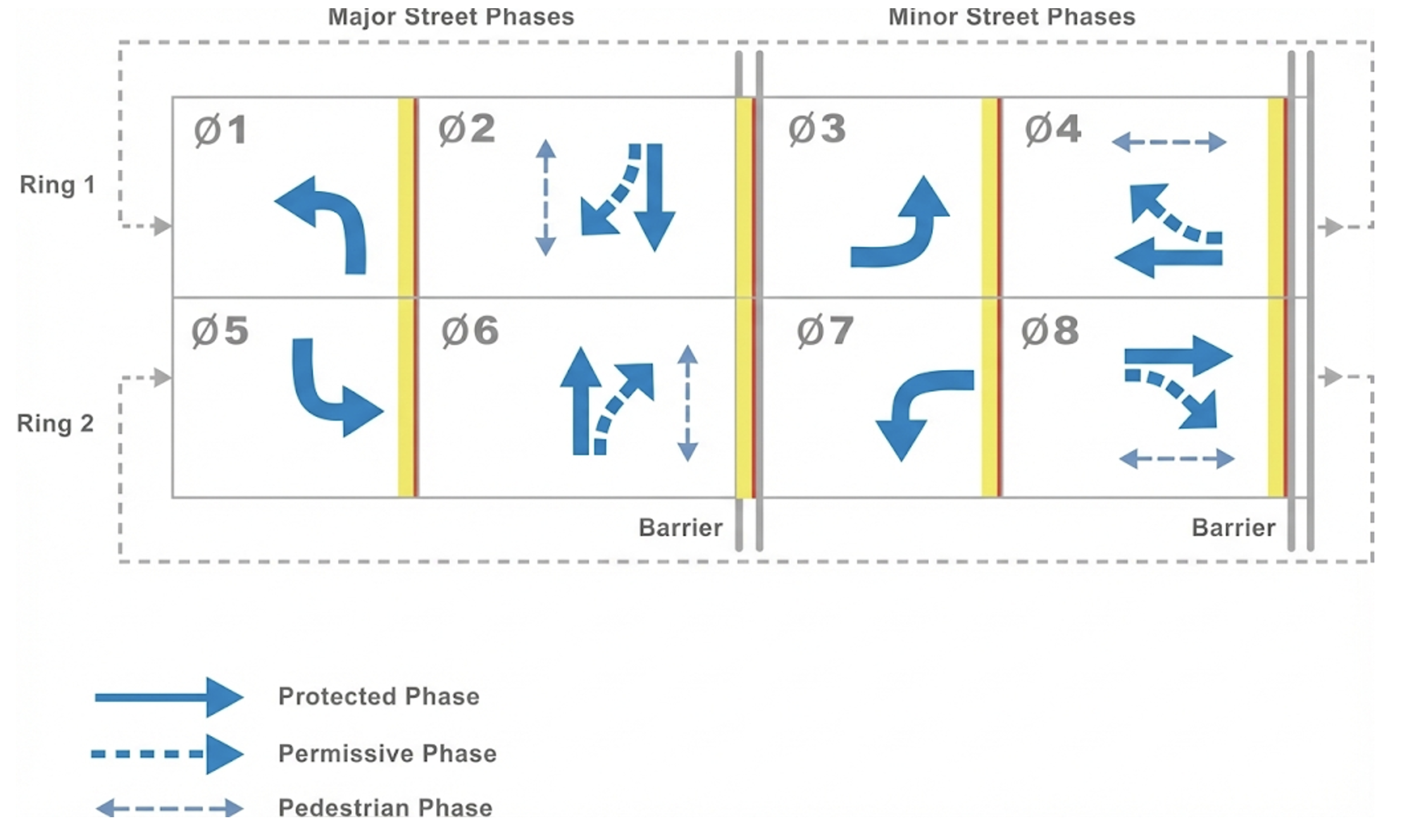}
    \caption{Ring-and-barrier diagram of a standard 4-leg intersection with 8 phases from Signal Timing Manual \cite{urbanik2015signal}.}
    \label{fig:ring-and-barrier}
\end{figure}

\noindent\textbf{Ring-and-barrier structure:}
A ring-and-barrier phase structure organizes phases into (typically) two rings that progress through phases subject to barrier constraints. The two rings may serve compatible phases concurrently, while barriers separate groups of phases that must be completed before advancing to the next group. The standard 2-ring 2-barrier structure with 8 phases from Signal Timing Manual \cite{urbanik2015signal} is shown in Fig \ref{fig:ring-and-barrier}. 

\noindent\textbf{Phase transition:}
A phase transition is the controller’s internal process of changing from the currently served phase pair to a newly commanded phase pair. This change is not instantaneous: after a phase-change command, the controller enforces clearance intervals (e.g., yellow change interval followed by all-red clearance) before the next phases become active (green).

\noindent\textbf{Real-time phase control:}
In the default mode of signal operation, the controller runs pre-defined timing plans and advances phases using its internal schedule and logic. In real-time phase control, ``real-time'' means that phase decisions are computed \emph{online} from the currently observed traffic state during simulation runtime (i.e., not pre-specified offline and not fixed for a time-of-day period). At each decision point, an external algorithmic entity (an agent) selects the next phase based on these live observations. The signal controller does not autonomously choose or progress phases; it follows and executes the agent’s commands as they are issued.
Specifically in this paper, the term of real-time refers to the TSC paradigm rather than to a hard real-time guarantee on every synchronization event between the signal controller and the external middleware, though in our experiments, the synchronization is empirically established.

\noindent\textbf{Signal and traffic states:} The signal state in this work is (narrowly) defined as the status (red, yellow, green) of each phase of an intersection. The traffic state excludes signal state information for clarity, which only represents the vehicular and roadway status such as queue length, wait time, etc. that can be retrieved from the simulation.

\noindent\textbf{NTCIP:}
National Transportation Communications for ITS Protocol is a family of ITS communication standards that specifies what transportation field devices expose for monitoring and control. Modern traffic signal controllers are standardized by the NTCIP 1202 standard.

\noindent\textbf{Middleware:}
The primary role of a middleware is ``to coordinate and enable communication between different layers or components while isolating much of the complexity of distribution into a single, well tested and well understood system abstraction.'' \cite{middleware-definition} Thus, a middleware is necessarily a unified interface between software and hardware.

\section{Hardware-in-the-Loop Simulation System}
We present the modules in our HILS in this section. We will first present the overall system architecture, then introduce the control and simulation modules in this section. We split out the middleware module in a separate section afterwards.

\noindent\textbf{Signal operating modes:} A preliminary system setup is that signal hardware controllers run on \emph{free mode} instead of the \emph{timing-plan mode} in our HILS, where no signal timing plans are programmed, and signals depend on phases with vehicle presence directly sending requests (i.e., phase calls) for service to alternate green lights. The software in our HILS is the only source that commands phase calls converted from control algorithm outputs.

\subsection{System Architecture}
\begin{figure}
    \centering
    \includegraphics[width=1\linewidth]{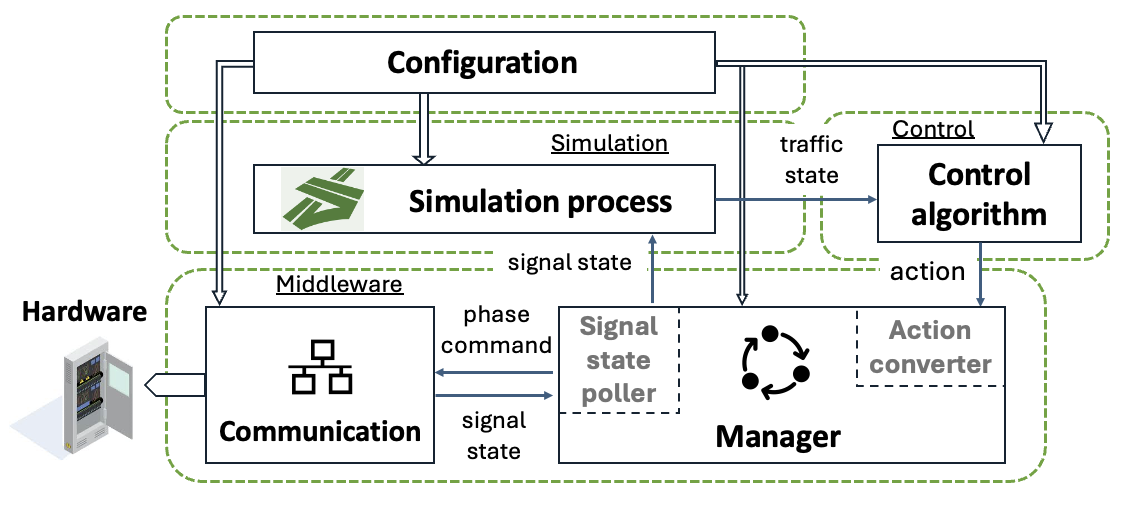}
    \caption{Diagram of the proposed HILS with a control-middleware-simulation architecture. The middleware consists of a manager and a communication layer. The manager commands control actions to the controller hardware and retrieves signal states from the controller via a communication layer.}
    \label{fig:architecture}
\end{figure}

The overall architecture of our proposed HILS is shown in Fig.~\ref{fig:architecture}. It is designed in a control-simulation-middleware structure. As traditionally designed, the control algorithm receives the traffic state and generates an action for the next timestep. The simulation module computes traffic state evolution. The newly introduced middleware serves as an orchestrator to interface with the traffic signal hardware and updates the signal states from the hardware controller.

Note that we refer \textit{actions} to the control algorithm's outputs, i.e., decisions, and \textit{phase commands} to the formatted message to be sent to the hardware controller as the post-processed form of actions.

\subsection{Control Module}
The control algorithm is integrated in the control module, which takes the traffic state from the simulation as input and outputs an action. Three major types of phase control action designs are supported in our HILS

Let $\mathcal{P}=\{p_n\}_{n\in N}$ denote the set of elementary \textit{phases}. We write $p_i \perp p_j$ if phases $p_i$ and $p_j$ are nonconflicting and may be served concurrently.
With concurrent service decisions restricted to a set of compatible \textit{phase pairs} $\mathcal{C}$, we define:
\begin{equation}\label{math:phase-pair}
    \mathcal{C}\triangleq \big\{\{p_i,p_j\}\subseteq \mathcal{P}:\; p_i \perp p_j \big\}.
\end{equation}
    
By nature of the ring-and-barrier design, all phases in the same ring are conflicting to each other. Thus, implicitly (\ref{math:phase-pair}) indicates $\mathcal{C}\subseteq \mathcal{P}_1\times\mathcal{P}_2$, where $\mathcal{P}_1$ and $\mathcal{P}_2$ are phases in ring 1 and ring 2.
Then the three phase control actions are:
\begin{itemize}
    \item \textbf{Phase selection:} the algorithm chooses the phases to serve for the next timestep as
    a compatible phase pair. The action space is the set of admissible pairs:
    \begin{equation}
        \mathcal{A}_{\text{sel}}=\mathcal{C}.
    \end{equation}
    \item \textbf{Phase switch:} when a fixed phase-pair sequence (as the two rings progress concurrently) is required, actions do not select
    arbitrary pairs. Let $(c^1,\dots,c^K)$ be a predefined sequence with $c^k\in\mathcal{C}$ and let
    $\sigma(c^k)=c^{(k+1)\mod k}$ (i.e., $k+1=1$ if $k=K$). The action is binary:
    \begin{equation}
        \mathcal{A}_{\text{sw}}=\{0,1\},
    \end{equation}
    where $a_t=0$ keeps the current pair ($c_{t+1}=c_t$) and $a_t=1$ advances to the next pair
    ($c_{t+1}=\sigma(c_t)$).
    \item \textbf{Phase duration:} the algorithm chooses the green duration of the next phase pair. Let $g_c^-$ and $g_c^+$ be the minimum and maximum green times associated with pair
    $c\in\mathcal{C}$. The action space is an interval mapped to the allowed green time conditioned on the next phase pair $c_t$:
    \begin{align} \label{math: phase duration}
        &\mathcal{A}_{\text{dur}}=[0,1] \mapsto [g_c^-, g_c^+].
    \end{align}
\end{itemize}

Note that phase selection is an acyclic action design, i.e., phase services are not in a fixed order. Phase switch and phase duration are cyclic as both require a predefined phase sequence to loop over. Our HILS is designed to support both acyclic and cyclic actions.

\subsection{Simulation Module}
We integrate SUMO~\cite{krajzewicz2012recent} as the traffic microsimulation software. Traditionally in software-only simulation for phase control, a phase action, including the phase transition stage, is directly commanded to the simulator which immediately yields the signal change. In our HILS, the simulation receives the phase states directly from the hardware controller via the middleware. Still, the road network geometry and traffic demands are configured in simulation files. The simulator computes how traffic evolves step by step given the current traffic state and the signal state.

\section{Middleware}
In timing-plan-based HILS, there is typically no explicit middleware as the simulation procedure is simply to upload a pre-computed signal-timing plan to the signal hardware before the simulation begins, without the need of real-time interaction.
However, real-time phase control requires a middleware to orchestrate (i.e., to manage and coordinate) recurring and frequent phase command and signal state retrieval processes. Our middleware design is organized into two modules by their functions: communication layer and state machine.

\subsection{Manager}\label{sec:state_machine} 
The manager module handles phase action commands and signal state retrievals and can be regarded as the commander for the real-time phase control implementation logic. It can command phases commands of any supported action design, track the phase transition progress, and handles errors. 

\subsubsection{Action Converter and Phase Lookup}
\label{sec: action converter}
We design an action converter submodule in the control module to convert any kind of action type to a unified action command to be transmitted to the communication layer. The unified action is a tuple $(p_{ring1}, p_{ring2})$, i.e., the chosen phase of each ring. This is coherent with the ring-and-barrier structure where both actions and can be translated from phase pairs. The manager internally runs a state model with the ring-and-barrier structure encoded, such that the phase sequence (if any) is programmed and can be looked up for following phases when a phase transition is called.

\subsubsection{Phase Conflict Detection}
Acyclic phase control design (phase selection) does not need a ring-and-barrier state model to look up new phases. How to guarantee the commanded phases are not conflicting in this case? Inspired by the mechanism of Malfunction Management Units (MMU) for signal controllers, we encode a 2D conflict matrix to detect phase pair conflicts. For a standard 4-leg intersection with 8 phases as in Fig. \ref{fig:ring-and-barrier}, the (symmetric) conflict matrix is:
\[
\mathbf{M}=
\begin{bmatrix}
0&0&0&0&1&1&0&0\\
0&0&0&0&1&1&0&0\\
0&0&0&0&0&0&1&1\\
0&0&0&0&0&0&1&1\\
1&1&0&0&0&0&0&0\\
1&1&0&0&0&0&0&0\\
0&0&1&1&0&0&0&0\\
0&0&1&1&0&0&0&0
\end{bmatrix},
\]
where the row index $i$ and column index $j$ corresponds to the two phases in a pair, and $M_{ij} = 1$ when a phase pair $[i,j]$ is conflicted.

\subsubsection{Issuing Phase Commands and Managing Phase Transition with a State Model} \label{sec: state model}
The core function of the manager is to issue phase commands to the controller. Once the control action is received and confirmed with no conflict, it is forwarded to the communication layer to issue a phase call. Meanwhile a thread is initiated to recurrently check if the signal state from the controller matches the commanded phases.

For convenience, we create an event-driven command state model for tracking the phase command and transition process with three states: \{\textsc{idle}, \textsc{on hold}, \textsc{timeout}\}.
\textsc{idle} state is the ready state where a phase command can be sent to execute at the communication layer. Once the phase command is sent, the manager jumps to \textsc{on hold} state which initiates a thread to recurrently verify phase-transition progress. During \textsc{on hold}, no new phase commands can be executed (as they will be dropped). If the phase transition is successful, i.e., the latest signal state matches the commanded phases in the recurrent verification, the state will revert to \textsc{idle}; otherwise, if the expected signal state is not observed within a time threshold $\tau_{\text{trans}}$, the system will fall in \textsc{timeout} state and pause the executing new phases until a manual recovery is called.

\subsubsection{Phase Duration Hold}

Phase duration actions introduce dynamic timestep lengths from the agent's perspective, and we describe here how they are realized in the HILS. Upon receiving a phase duration action, the manager looks up the minimum and maximum green time of the upcoming phase from the configuration file and computes the actual phase duration according to Eq.~\ref{math: phase duration}, for which a countdown timer is initialized. The phase transition then executes while the manager remains in the \textsc{on hold} state. Once the transition completes, the \textsc{on hold} state is extended until the timer expires, at which point the prescribed phase duration has been fully served. The manager then reverts to \textsc{idle}, and the control agent is invoked again to observe the current state and generate the duration action for the next phase.

\subsubsection{Polling Signal States}
The simulation and the manager need up-to-date signal states to calculate new steps and check if the transition is over. Having each part separately polling signal states from the controller unnecessarily adds communication traffic. Therefore, we use a daemon thread (i.e., a background thread that continuously runs without blocking the main program from exiting) which polls the controller at a configurable interval. This thread is independent of the simulation and the state machine, and the latest signal states are pushed to a shared cache which other processes and threads can read. This method alleviates the communicate workload and supports signal state update at a high frequency. The poller tracks consecutive communication failures: after a configurable number $n_{\text{timeout}}$ of consecutive timeouts $t$, it raises the \textit{communication failure} flag and sets the manager to \textsc{timeout} state.

\subsubsection{Error Detection and Recovery} \label{sec:error}
Three kinds of errors are handled. First, the communication is problematic that no response is received from the controller. Second, signal transition is not successful such that the state machine stays in \textsc{on hold} state longer a time threshold $\tau_{\text{trans}}$. Third, simulation time drifts (i.e., a simulation step calculation takes longer than the step length) for $n_{\text{drift}}$ times consecutively, such that the simulation time and wall clock time lose synchronization.

When any of these errors happens, a system-wide \textit{timeout event} flag will be set and the state machine will jump to \textsc{timeout} state. Communication for phase commands and signal state polling will be both halted. A manual recovery function is available in the system for external processes such as the control algorithm to call, and once called, a recovery procedure initiates to re-read the controller's current states, reset the timeout flag, and reset the state machine to \textsc{idle} state.

\subsection{Communication Layer}\label{sec:comm_layer}
The communication layer functions as (1) sending phase commands to the hardware controller, and (2) polling signal states back from the hardware controller. The two basic functions in NTCIP communication are used: SET and GET. Note that calling both functions will receive a controller response if the communication is successful. 

\subsubsection{Sending Phase Commands}
\label{sec: sending command}
To send a phase command, the communication layer is called with a function by the manager to issue a phase call to specific phases. The communication layer packs the phase call message into a packet and sends to the hardware with NTCIP SET function.

\subsubsection{Receiving Signal States}
Receiving signal state function is called by the poller. The communication layer sends a series of NTCIP GET functions to respectively acquire the red, yellow, and green phases.

\subsection{Phase Command Workflow} \label{sec: workflow}
\begin{figure}
    \centering
    \includegraphics[width=1\linewidth]{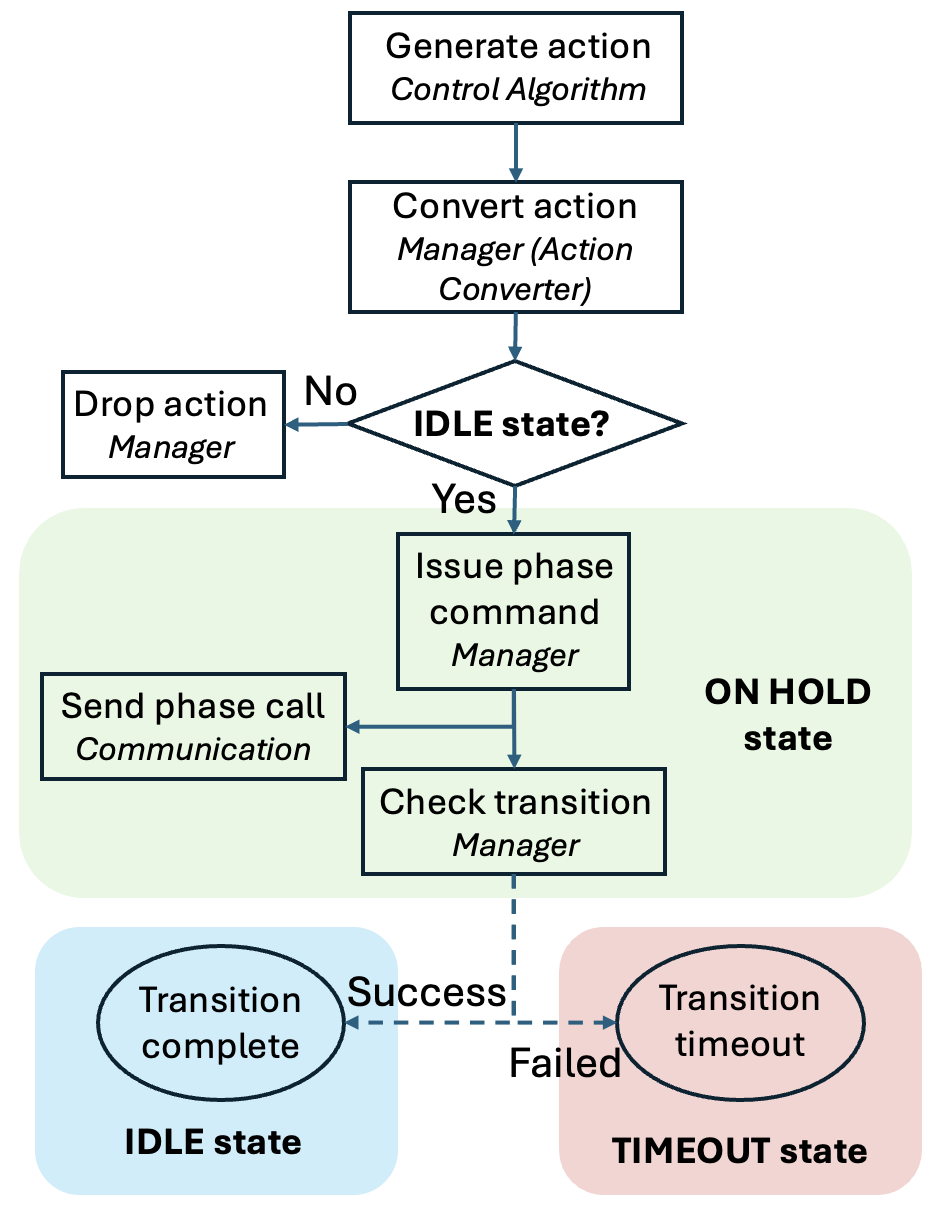}
    \caption{Flow diagram of the process of issuing a phase command. Each rectangle box is an event happened in a \textit{module} (italic font). The phase transition terminates either completed or timed out as marked by the ellipses. The three colored areas indicate which Manager transition state (discussed in Sec. \ref{sec: state model}) the covered events are at.}
    \label{fig:phase-command}
\end{figure}

The complete life cycle of a phase command proceeds as follows. The control algorithm generates an action and sends it to the manager, which first invokes the action converter to transform it into a unified form (Sec.~\ref{sec: action converter}); phase conflict detection is incorporated into this step for simplicity. The manager then checks whether its state is \textsc{idle}: if not, the action is dropped; otherwise, the manager transitions to \textsc{on hold} and issues the phase command, simultaneously triggering (1) the communication layer to transmit the phase call (Sec.~\ref{sec: sending command}), and (2) a monitoring thread to verify completion of the phase transition (Sec.~\ref{sec: state model}). Upon confirmation of a successful transition, the manager reverts to \textsc{idle}. Upon any failure, it transitions to \textsc{timeout}, where the error detection, handling, and recovery mechanisms are discussed in Sec.~\ref{sec:error}.

\section{Experiments}
The experiments aim to validate the controller-facing functionality and timing behavior of the proposed HILS. The HILS setup is presented first. Then we show a slice of phase transition trajectory to demonstrate the temporal sequence of phase transitions with a focus on the start and end of a transition. Then we show the system error handling behaviors by simulating predefined errors in test scripts. Finally, we study the communication latency in three types of connection setups.

\subsection{HILS Setup}
In following experiments, the HILS is connected to a signal cabinet in a lab setup (without traffic lights) with an Econolite Cobalt controller produced in 2021, as shown in Fig. \ref{fig:cabinet}. The SUMO simulation scenario is one four-leg intersection from 1000DaySim \cite{zhang20251000daysim}. For convenience, all phases share the same parameters: 3\,s minimum green, 20\,s maximum green, 3\,s yellow transition, and 2\,s red clearance. Only four phase pairs are considered: $[1,5],[2,6],[3,7],[4,8]$. Parameter settings are as follows: $\tau_{\text{lock}}=5\text{ s}$, $\tau_{\text{udp}}=1\text{ s}$, $\tau_{\text{trans}}=10\text{ s}$, $n_{\text{timeout}}=5$, $n_{\text{drift}}=5$. The \texttt{NtcipPoller} polling frequency is 10\,Hz, and SUMO step length is 0.25\,s. Phase control interval for phase selection and phase switch is 10\,s (phase duration does not have uniform control interval). The HILS code is hosted on a 2021 MacBook Pro with an M1 Pro Chip and 32\,GB RAM. The main programming language is Python v3.12.4. 

\subsection{Phase Transition Trajectory}

\begin{figure}
    \centering
    \begin{subfigure}{\linewidth}
        \centering
        \includegraphics[width=\linewidth]{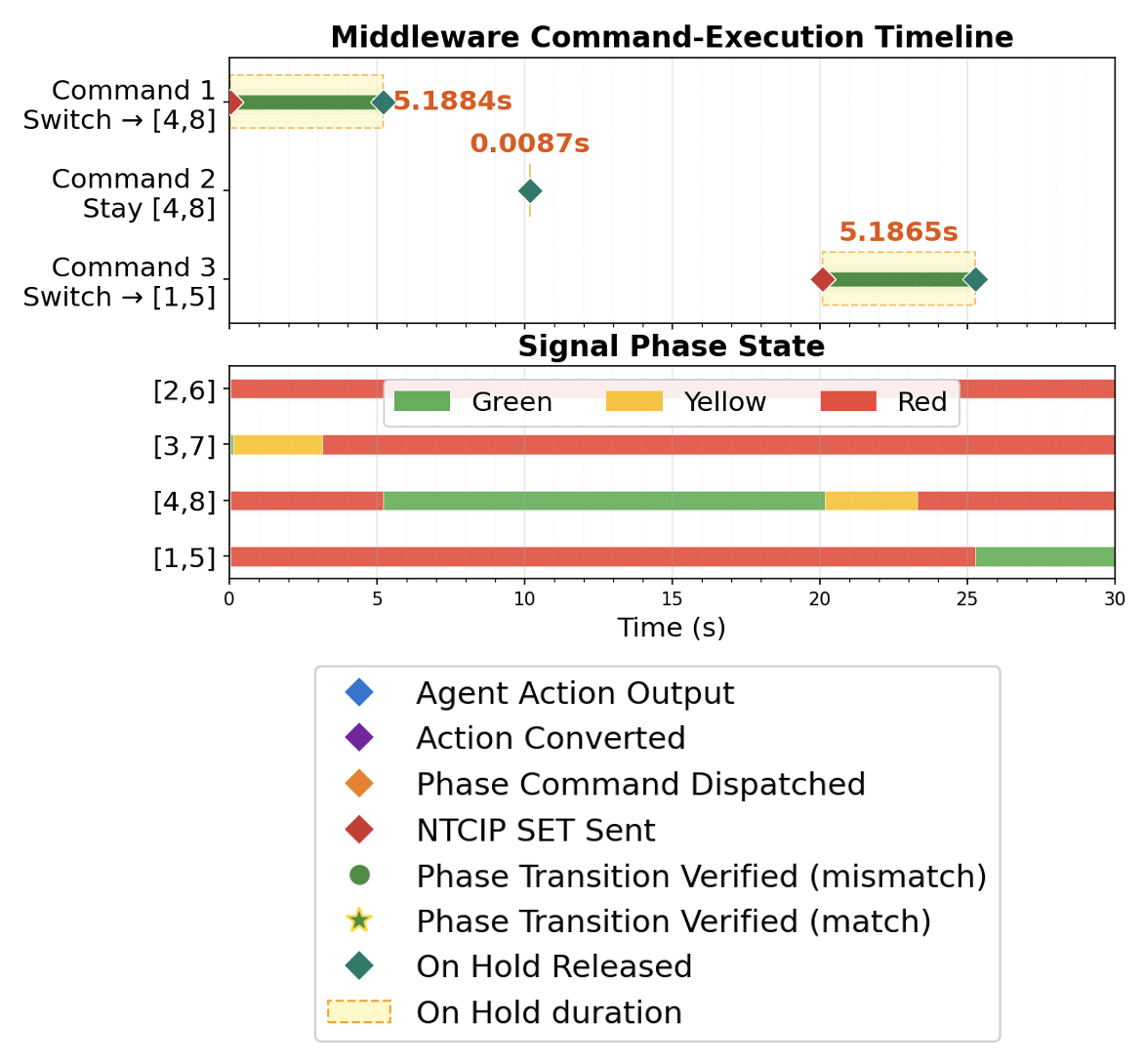}
        \caption{Phase transition trajectory from three phase switch commands. Up subfigure: phase command related event timeline (corresponding to Sec. \ref{sec: workflow}). Middle subfigure: signal state trajectory for phase pairs. Bottom: legend of phase command related events shared by both figures.}
        \label{fig:trajectory-main}
    \end{subfigure}

    \begin{subfigure}{\linewidth}
        \centering
        \includegraphics[width=\linewidth]{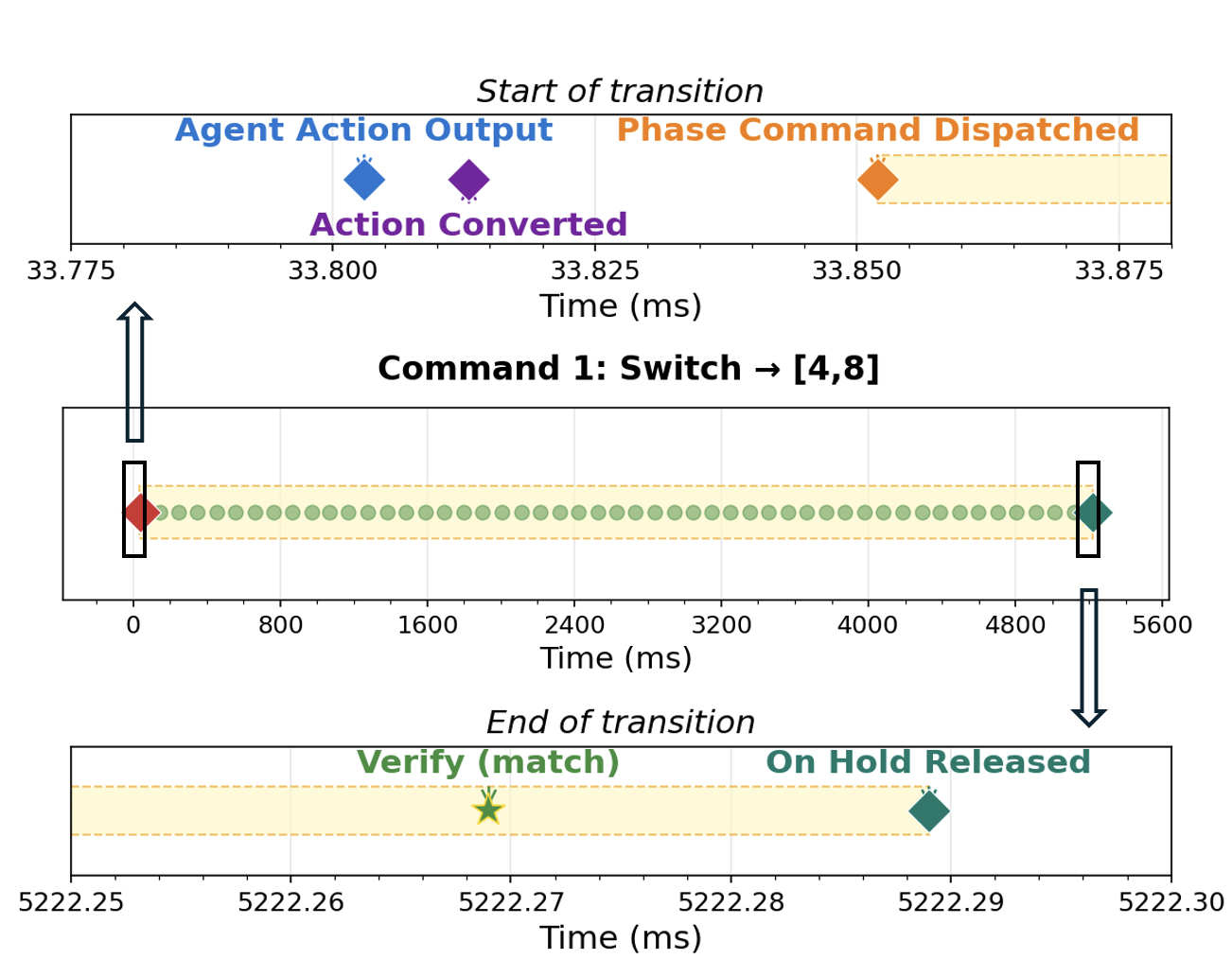}
        \caption{Zoom-in transition trajectory for Command 1. Center: the full transition for Command 1. Up: The start of transition from agent generating an output to phase command being dispatched to the communication layer. Bottom: The end of transition from a matched transition verification to the \textsc{on hold} state being reverted back to \textsc{idle}.}
        \label{fig:trajectory-zoom}
    \end{subfigure}
    \caption{Phase transition trajectory from three phase switch commands and a zoom-in view of Command 1. Events share the same legend.}
    \label{fig:trajectory}
\end{figure}

Fig.~\ref{fig:trajectory} visualizes a slice of the phase transition trajectory from the system log and contains two subfigures. Fig.~\ref{fig:trajectory-main} illustrates the command execution process of three consecutive commands. The signal initially serves phase $[3,7]$. At the first timestep ($t=0\text{s}$), Command~1 requests a switch to phase $[4,8]$. Command~2, issued at the second timestep, requests to remain at $[4,8]$. Command~3 then issues a phase switch to $[1,5]$. The phase trajectory is reconstructed from the signal state logged via the shared cache of the signal state poller. Upon receiving a new phase call, the signal enters a yellow transition stage followed by a red clearance period before beginning to serve the new phase.

Fig.~\ref{fig:trajectory-zoom} zooms into the transition process triggered by Command~1, with the start and end events further magnified due to their narrow time windows. As shown in the upper subfigure, the action output by the control agent is converted and dispatched as a command by the manager; the dispatch event simultaneously triggers the \textsc{on hold} state and initiates the recurrent transition verification thread. These three events occur within a 0.05\,ms window. The NTCIP SET function, which requires a response from the controller, completes 7.494\,ms after the dispatch event and is therefore omitted from the upper subfigure. The verification thread subsequently polls for transition completion every 0.1\,s, producing the sequence of green circles in the middle subfigure. Once a matching result is returned, the \textsc{on hold} state is released and the manager reverts to \textsc{idle}, as shown in the bottom subfigure.

The hold duration per transition ranges from 5.18\,s to 5.20\,s in our records, intentionally exceeding the nominal transition time of 5\,s for reliability. The dominant source of this overhead is the 0.1\,s signal state polling interval.

\subsection{Middleware Internal Latency}

We define the \textit{middleware internal latency} as the elapsed time from the moment the control agent outputs an action to the moment the NTCIP SET command is dispatched (exclusive), capturing the overhead introduced by the internal phase handling pipeline of the middleware.
Latency measurements across all three action types are summarized in Table~\ref{tab:latency}.

\begin{table}[h]
\centering
\caption{Middleware internal latency by action type. N: number of samples.}
\label{tab:latency}
\begin{tabular}{lccc}
\toprule
Action Type & $N$ & Mean (ms) & Std (ms) \\
\midrule
\texttt{phase\_selection} & 71 & 0.6464 & 0.6278 \\
\texttt{phase\_switch}    & 91 & 0.8745 & 1.0798 \\
\texttt{phase\_duration}  & 54 & 0.6906 & 0.6729 \\
\bottomrule
\end{tabular}
\end{table}

All three action types exhibit sub-millisecond mean latency, confirming that the middleware imposes negligible computational overhead on the control loop. \texttt{phase\_switch} incurs the highest mean latency (0.8745\,ms) and the largest variance (std: 1.0798\,ms). \texttt{phase\_selection} and \texttt{phase\_duration} show comparable and lower mean latencies, with relatively tighter distributions. The elevated standard deviations across all types suggest occasional outlier events, possibly attributable to thread scheduling jitter or transient system load, rather than systematic processing cost. Overall, the middleware latency is well within acceptable bounds for real-time TSC. Nevertheless, we note that these measurements should not be interpreted as full end-to-end real-time synchronization guarantees, because they exclude communication response time, controller-internal clearance behavior, and transition verification.

\section*{Conclusion}
In this paper, we propose the first Hardware-in-the-Loop Simulation (HILS) testbed supporting real-time traffic signal phase control. Different from conventional HILS, our HILS enables external control algorithm to generate and command signal phases in real time, which matches advanced Traffic Signal Control (TSC) formulations. To address the real-time control challenges, the core of our HILS is a middleware module which incorporates a manager and a communication layer to safely interface software agents with hardware controllers. Specifically, the manager conducts a series of events to convert a control action to a phase command, regularly polls the signal states and share them with other parts of the HILS with a shared cache, and handles errors and conflicts. The communication layer establishes communication with the hardware controller which is compliant to NTCIP standard.
We demonstrated the full execution sequence of phase commands and transitions experimentally, and analyze the middleware-induced action execution delay at sub-millisecond level. 

This demonstration of real-time phase control realization bridges the gap between advanced TSC design and hardware realization. This testbed paves the way for scalable, real-world deployments of advanced traffic signal control strategies. Our next step is to extend this framework to multi-controller settings and improve the error handling design.

\bibliographystyle{IEEEtran}
\bibliography{references}

\end{document}